\pdfoutput=1
\documentclass[sigconf]{acmart}
\AtBeginDocument{%
  }
\setcopyright{acmlicensed}
\usepackage{times} 
\usepackage{latexsym}
\usepackage[T1]{fontenc}
\usepackage[utf8]{inputenc}
\usepackage{microtype}
\usepackage{graphicx}
\usepackage{algorithmic}
\usepackage{textcomp}
\usepackage{xspace}
\usepackage[ruled, lined, linesnumbered, commentsnumbered, longend]{algorithm2e}
\usepackage{multirow,multicol}
\usepackage{colortbl}
\usepackage{caption}
\usepackage{subfigure}
\usepackage{tcolorbox}
\usepackage{mathtools}
\usepackage{tikz}
\usepackage{units}
\usepackage{listings}
\usepackage{ulem}
\usepackage{bbding}
\usepackage{bm}
\definecolor{backcolour}{rgb}{0.95,0.95,0.92}
\usepackage{enumitem}

\SetCommentSty{mycommfont}

\newcommand*\circled[1]{\tikz[baseline=(char.base)]{
            \node[shape=circle,fill,inner sep=1pt,font=\footnotesize] (char) {\textcolor{white}{#1}};}}

\usepackage{tabularx}
\usepackage{listings}
\usepackage{color}

\definecolor{dkgreen}{rgb}{0,0.6,0}
\definecolor{gray}{rgb}{0.5,0.5,0.5}
\definecolor{mauve}{rgb}{0.58,0,0.82}

\usepackage{xcolor}

\graphicspath{ {./figures/} }
\usepackage{textgreek}
\usepackage{float}

\newcommand{\ourtool}{\textsc{SliceT5}\xspace}

\AtBeginDocument{%
  }
\begin{document}

\title{\ourtool: Static Program Slicing using Language Models with Copy Mechanism and Constrained Decoding}

\author{Pengfei He}
\affiliation{%
  \institution{University of Manitoba}
  \city{Winnipeg}
  \country{Canada}
}
\email{hep2@myumanitoba.ca}

\author{Shaowei Wang}
\affiliation{%
  \institution{University of Manitoba}
  \city{Winnipeg}
  \country{Canada}
}
\email{shaowei.wang@umanitoba.ca}

\author{Tse-Hsun Chen}
\affiliation{%
  \institution{Concordia University}
  \city{Montreal}
  \country{Canada}
}
\email{peterc@encs.concordia.ca}




\begin{abstract}
Static program slicing is a fundamental technique in software engineering. Traditional static slicing tools rely on parsing complete source code, which limits their applicability to real-world scenarios where code snippets are incomplete or unparsable. While recent research developed learning-based approaches to predict slices, they face critical challenges: (1) Inaccurate dependency identification, where models fail to precisely capture data and control dependencies between code elements; and (2) Unconstrained generation, where models produce slices with extraneous or hallucinated tokens not present in the input, violating the structural integrity of slices.
To address these challenges, we propose \ourtool, a novel slicing framework that reformulates static program slicing as a sequence-to-sequence task using lightweight language models (e.g., CodeT5+). Our approach incorporates two key innovations. First, we introduce a copy mechanism that enables the model to more accurately capture inter-element dependencies and directly copy relevant tokens from the input, improving both dependency reasoning and generation constraint. Second, we design a constrained decoding process with (a) lexical constraint, restricting outputs to input tokens only, and (b) syntactic constraint, leveraging Tree Similarity of Edit Distance (TSED) monotonicity to detect structurally invalid outputs and discard them.
We evaluate \ourtool on CodeNet and LeetCode datasets and show it consistently outperforms state-of-the-art baselines, improving ExactMatch scores by up to 27\%. Furthermore, \ourtool demonstrates strong performance on incomplete code, highlighting its robustness and practical utility in real-world development environments.

\end{abstract}



\maketitle


\section{Introduction}\label{sec:intro}
Static program slicing plays a crucial role in software engineering tasks such as vulnerability analysis \cite{vdpm,vdp} and debugging~\cite{slice1984,xu2005brief,slicevd}. 
Compared to dynamic slicing, which requires executing the program to capture runtime behavior~\cite{harman2001overview,agrawal1990dynamic}, static slicing does not depend on a runtime environment or specific test code without execution~\cite{binkley2007empirical}, enabling it more practical and broadly applicable~\cite{acharya2011practical,xu2005brief}.

Conventional static slicing tools, such as JavaSlicer \cite{javaslicer,javaslicerpaper} and CPP-Slicer \cite{cppslicer}, typically parse the Abstract Syntax Tree (AST), and insert data dependency edges to construct a System Dependence Graph (SDG). Once the SDG is built, slicing can be performed as a graph reachability problem by selecting the node corresponding to the slicing criterion and traversing the graph to identify all parts of the program that may affect or be affected by that criterion~\cite{javaslicerpaper}. However, a prerequisite of these traditional approaches is fully compilable source code, often at a minimum of method-level granularity~\cite{learning}. This limits their applicability to real-world scenarios, where code snippets, such as those found on online forums like Stack Overflow, are frequently incomplete and unparsable~\cite{unparsable}, despite being rich sources of developer knowledge.

Recently, there has been growing interest in leveraging language models (LMs) to automatically predict static program slices. Yadavally et al. \cite{learning} fine-tuned CodeBERT \cite{codebert} and GraphCodeBERT~\cite{graphcodebert} to predict relevant slices by computing the similarity between individual statements and the slicing criterion. However, this method operates at the statement level with limited contextual awareness and cannot feasibly incorporate all statements within a method, leading to a lack of inter-statement dependency modeling. Shahandashti et al. \cite{llmslicer} utilized advanced foundation models (FMs) such as GPT-4 \cite{gpt4} and Gemma~\cite{gemma2}, combined with prompting techniques such as retrieval-augmented generation (RAG) \cite{allyouneed} and chain-of-thought reasoning (COT) \cite{cot} to predict program slices. However, the effectiveness of these models on static program slicing remains limited. They often suffer from hallucinations due to insufficient logical reasoning over complex program \cite{llmslicer}. Additionally, foundation models are expensive to deploy in practice, often incurring costs through API calls and are heavy on hardware requirements.

To address these limitations, we formulate static program slicing as a sequence-to-sequence (seq2seq) problem and propose a novel slicing approach built on lightweight LMs (i.e., CodeT5~\cite{codet5} and CodeT5+~\cite{codet5+}). However, directly fine-tuning off-the-shelf LMs as slicers can be suboptimal. In our investigation (Sec.~\ref{sec:motivation}), we observe two major challenges: \circled{1} \textbf{Inaccurate dependency identification}. An effective slicer requires precise identification of relationships between code elements (e.g., data and control dependencies) for a given slicing criterion. Language models currently struggle with this crucial aspect, which often leads to missing or including extra elements. \circled{2} \textbf{Unconstrained generation}. The generated slice must be an exact subpart of the original code snippet, with all tokens precisely extracted, and their order and structure preserved. However, current models often exhibit unconstrained generation, i.e., producing code elements not present in the original code snippet.

To address the challenges, we propose a novel approach, namely \ourtool, which integrates \textbf{copy mechanism} and \textbf{constrained decoding} to generate more accurate slices. Specifically, we improve the existing language models in two folders. 
First, we integrate a copy mechanism into the model. The copy mechanism strengthens the model's ability to learn and capture relationships between code elements, directly addressing Challenge \circled{1}. Additionally, it provides a direct pathway for transferring exact tokens from the input to the output, which helps mitigate Challenge \circled{2}. At each decoding step, the model calculates a copy probability, determining when to directly copy tokens from the input sequence.
Second, we develop a novel constrained decoding process by integrating two constraints to further tackle the Challenge \circled{2}. \textbf{Lexical constraint}: We restrict the decoding process to only accept tokens present in the original code, preventing the generation of invalid tokens.
\textbf{Syntactic constraint}: We measure the syntactic alignment between the generated slice and the original code using Tree Similarity of Edit Distance (TSED)~\cite{tsed}. If the TSED score of the generated code does not monotonically increase as expected, it is a sign that the code is structurally biased (e.g., through repetition or misalignment). Such candidates are discarded to ensure only syntactically coherent outputs are retained.

We evaluated \ourtool on two datasets CodeNet and Leetcode. \ourtool consistently outperforms SOTA baselines (e.g., FM-based slicer~\cite{llmslicer} and NS-slicer~\cite{learning}) across different datasets and evaluation metrics. For instance, \ourtool outperforms the best NS-slicer by at least 6.4\% and 27\% in terms of
ExactMatch, on the two datasets, respectively. We also evaluated \ourtool on incomplete code snippets, and \ourtool still consistently outperforms all SOTA baselines, demonstrating the robustness of \ourtool for unparsable code.

We make the following contributions:
\begin{itemize}
    \item We propose the first end-to-end solution for static program slicing by improving the existing language model's architecture and developing novel constrained decoding.
    \item Through extensive evaluation, we demonstrated that our approach achieves significant performance improvements over the SOTA baselines. 
    \item We make our code public \footnote{https://anonymous.4open.science/r/staticsliceT5-4E22}.
\end{itemize}

\section{Problem Formulation and Challenges}\label{sec:overview}
In this section, we first present the formal definition of static program slicing. We then highlight the challenges that arise when directly fine-tuning language models for slicing, which motivates the design of our proposed method.

\subsection{Problem Formulation}\label{sec: problem}

\begin{figure}[!htbp]

\begin{tcolorbox}[boxsep=1pt,left=1pt,right=1pt,top=1pt,bottom=1pt]
\begin{lstlisting}[language=Java,frame=single,framerule=0pt,basicstyle=\ttfamily\footnotesize]
// Example (1) - Inaccurate dependency identification
// Expected slice:
7: int temp
8: if(C <= A){
12: temp = B;
...
// Generated slice:
7: int temp
8: if(C <= A){
9: (*@\textcolor{red}{temp = A;}@*)
10: (*@\textcolor{red}{A = C;}@*)
12: temp = B;
...
\end{lstlisting}
\end{tcolorbox}

\begin{tcolorbox}
[boxsep=1pt,left=1pt,right=1pt,top=1pt,bottom=1pt]
\begin{lstlisting}[language=Java,frame=single,framerule=0pt]
// Example (2) - Replacing a variable with non-existent one
// Expected slice:
...
 7: int cnt = 0;
 10:for(int i=cnt;i>=0;i--) {
 11:if(i>0) {long y = x[i];
 12:long Codepoint = 97+y};

// Generated slice:
...
 7: int cnt = 0;
 10:for(int i=cnt;i>=0;i--) {
 11:if(i>0) {long y = x[i];
 12:long (*@\textcolor{red}{keta}@*) = 97+y};
\end{lstlisting}
\end{tcolorbox}

% ...

% ...

\begin{tcolorbox}
[boxsep=1pt,left=1pt,right=1pt,top=1pt,bottom=1pt]
\begin{lstlisting}[language=Java,frame=single,framerule=0pt]
// Example (3) - Generating non-exitent statement
// Expected slice:
4: int one = 0, five = 0, ten = n;
6: try {
8:      if (one* 1 + five * 5 + ten * 10 >y)
...
// Generated slice:
4: int one = 0, five = 0, ten = n;
6: try {
8:      if (one * 1 + five * 5 + ten * 10 >  y (*@\textcolor{red}{* 10 * y * y * y * z * ten * 10 * y * z * ten * 10 * y * z}@*)
...
\end{lstlisting}
\end{tcolorbox}
\caption{
Motivating examples of three wrong program slices produced by directly fine-tuned CodeT5+.
}
\label{fig:motivation}
\end{figure}


We formulate \textbf{static program slicing} as a sequence-to-sequence code transformation task.
Formally, the input code snippet to be sliced is represented as:
\begin{align}
\bm{x} = \{s_1, s_2, \dots, s_i, \dots, s_N; v; [n]\}
\end{align}
where \( s_i \) denotes a statement of the code snippet $\{s_1, s_2, \dots, s_N\}$, \(v\) is the \textbf{variable of interest}, \([n]\) is the line number of \(v\). Each statement \( s_i \) is further represented as a sequence of code tokens \( \bm{s_i} = \{ t^{(i)}_1 , t^{(i)}_2 , \dots , t^{(i)}_{M_i}\} \).
 
Given $\bm{x}$, the slicer $P(\bm{y}|\bm{x})$ should predict a slice $\bm{y}$, where $\bm{y} \subseteq \bm{x}$ comprises the minimal set of statements, known as the backward slice, that \textbf{semantically influence} the value of the slicing criterion $v$. Formally, $\bm{y}$ is defined as:
\begin{align}
    \bm{y} = \{s_{i_1}, s_{i_2}, \dots, s_{i_k}\} \subseteq \{s_1, s_2, \dots, s_{[n]}\}, \quad i_1 < i_2 < \dots < i_k
\end{align}
The output slice $\bm{y}$ is expected to satisfy the following properties:
\begin{itemize}
  \item \textbf{Accuracy}: the generated slice, \(\bm{y}\), should include all and only the relevant statements that are expected to appear in the correct program slice with respect to the slicing criterion.
  \item \textbf{Code element preservation}: the slice \(\bm{y}\) must be a subpart of the original code snippet \(\bm{x} = \{s_1, s_2, \dots, s_N\}\), meaning that every statement \(s_i \in \bm{y}\) must be exactly extracted from \(\bm{x}\) without modification, meanwhile preserving the same ordering. At the lexical level, each token \(t^{(i)}_j\) in the selected statements must appear verbatim in the original code and preserve the same ordering as in the code snippet.
\end{itemize}

\subsection{Limitations of Vanilla Fine-Tuned
Language Models for Program Slicing}\label{sec:motivation} 
A seq2seq model is a deep learning architecture designed to convert an input sequence into an output sequence~\cite{sutskever2014sequence}. Transformer-based seq2seq models are particularly preferred due to their strong performance and versatility, and have been adapted for programming tasks, most notably in models such as CodeT5~\cite{codet5} and CodeT5+\cite{codet5+}. These models have achieved state-of-the-art results in code understanding and generation tasks, and are easily fine-tuned for a wide range of downstream software engineering applications, including type inference~\cite{typeT5} and code compression~\cite{codeprompt}.
Intuitively, seq2seq models can also be employed to learn the dependencies between code elements, making them a natural fit for the task of program slicing. More formally, the goal is to model the conditional probability $P(\bm{y}|\bm{x})$, where $\bm{y}$ denotes the sequence of program statements that constitute the desired slice, and $\bm{x}$ is the input program along with the slicing criterion. However, generating accurate and element-preserving slices remains challenging when relying solely on direct fine-tuning of off-the-shelf models. In our empirical study, we identify several key limitations of directly fine-tuned models (e.g., CodeT5 and CodeT5+) when applied to the static program slicing task:

\begin{itemize}[leftmargin=0.in]
   
    \item[] \circled{1} \textbf{Inaccurate dependency identification.} It is challenging for language models (LMs) to accurately capture the dependencies between code elements. As a result, they often miss relevant statements or include irrelevant ones when performing slicing. For instance, in the Example (1) shown in Figure~\ref{fig:motivation}, the generated slice mistakenly includes additional statements, line 9 (\texttt{temp = A;}) and line 10 (\texttt{A = C;}), which are not required. Rather than identifying true data dependencies, the model tends to rely on surface-level patterns or positional proximity, leading to incorrect inclusion of unrelated statements in the slice.
    
    \item[] \circled{2} \textbf{Unconstrained generation.} The model exhibits unconstrained generation, i.e., producing code elements that are not presented in the original code. For instance, language models can struggle with rare tokens, often replacing them with semantically or syntactically incorrect alternatives, such as generating an frequent alternative \texttt{keta} instead of the correct identifier \texttt{codepoint} in Example (2) in Figure~\ref{fig:motivation}, which leads to a syntactic error. In fact, LM-generated code is often prone to lexical errors \cite{errors}. In addition, models tend to generate unrelated logic or duplicating statements, such as Example (3) in Figure~\ref{fig:motivation}, where the generated statement \texttt{y * 10 * y * y * y * z * ten ...} is not the exact statements presented in the original code, and the sub statement \texttt{z * ten * 10 * y} repeats. These errors collectively underscore the limitations of the vanilla fine-tuned model in producing accurate and faithful slices. 
\end{itemize}

\section{Methodology}\label{sec:framework}

\begin{figure}
    \centering
    \includegraphics[width=0.9\linewidth]{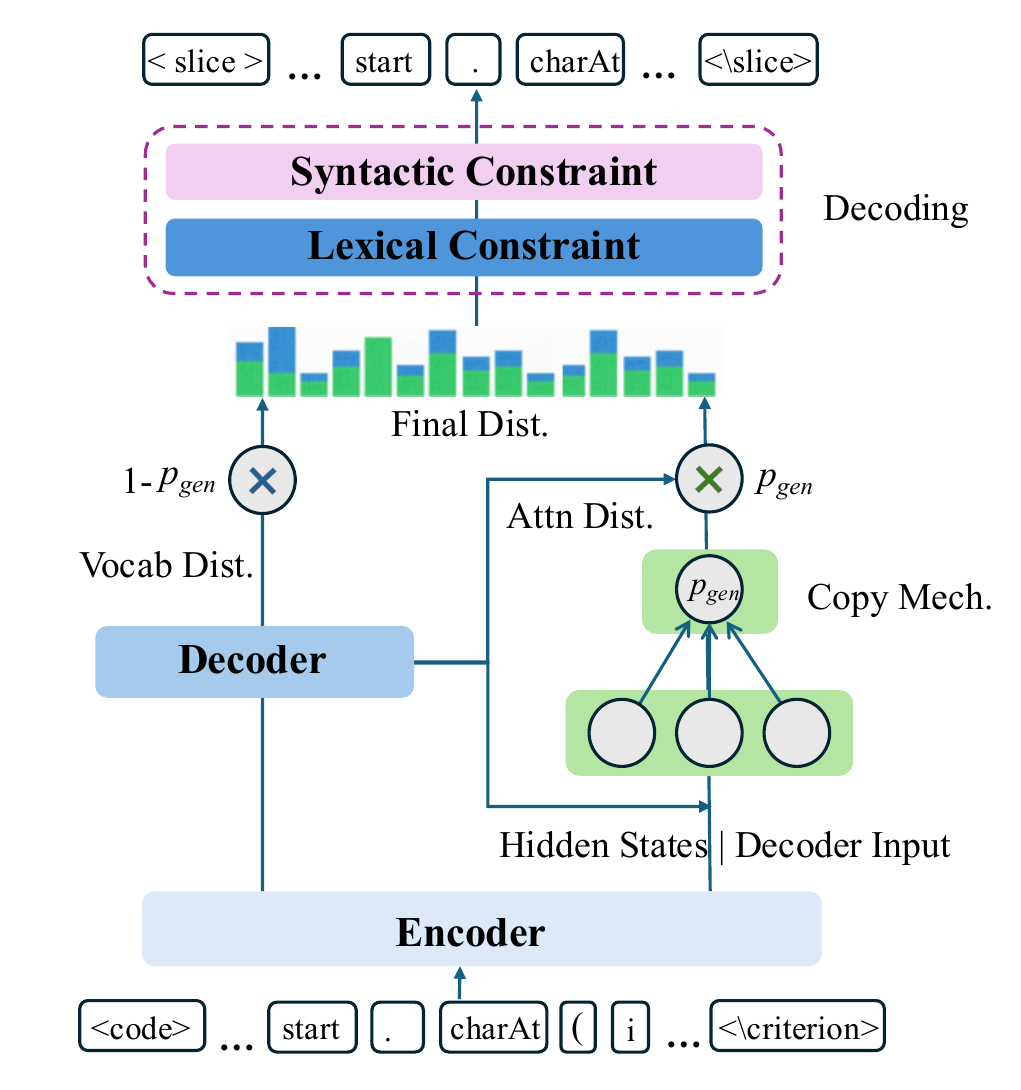}
    \caption{An overview of \ourtool.}
    \label{fig:framework}
    \vspace{-0.3cm}
\end{figure}


To address the limitations observed in Section~\ref{sec:motivation}, we propose a new design of the static program slicing framework, \ourtool, which integrates copy mechanism and constrained decoding to generate more accurate and element-preserved slices. Figure~\ref{fig:framework} presents the overall framework of \ourtool. More specifically, we improve the original Transformer-based models in two ways:

\noindent\textbf{Architecture.} We first integrate a copy mechanism~\cite{copy} into the model. The copy mechanism allows the model to directly copy tokens from the input sequence to the output, rather than always generating output tokens from the fixed vocabulary. The copy mechanism addresses the limitations in two ways. First, the copy mechanism enhances the model to capture the relevant statements based on the given slicing criterion (to address \circled{1}). Second, the code mechanism provides a direct pathway for transferring exact tokens from the input to the output (to address \circled{2}). At each decoding step, the model computes a copy probability that governs whether the relevant tokens should be copied directly from the input sequence. This mechanism enables the model to better handle rare identifiers by circumventing vocabulary limitations and enhancing the fidelity of reproducing rare tokens, while increasing the likelihood of sampling tokens from the original code for output.

\noindent\textbf{Decoding.} To mitigate the limitation \circled{2}, we enforce decoding process of beam search by incorporating lexical and syntactic constraints. 
\textbf{Lexical constraint}: We restrict the decoding process to only accept tokens present in the original code, preventing the generation of invalid tokens that does not present in the original code. 
\textbf{Syntactic constraint}: We evaluate the syntactic alignment between the generated slice and the original code using Tree Similarity of Edit Distance (TSED)~\cite{tsed}. Since a valid program slice is a subsequence of the input code snippet, the TSED score of the partially generated sequence is expected to increase monotonically during decoding as more code elements are generated correctly. Conversely, if the generated code deviates structurally, due to syntactic inconsistencies such as repetition or misalignment, the TSED score may drop, indicating a violation of structural coherence. In such cases, the generation is considered invalid, and the output is discarded to ensure that only syntactically well-formed slices are retained.

We elaborate on each mechanism in detail below.

\subsection{Copy Mechanism}\label{sec:copy}

The copy mechanism, originally proposed in pointer networks \cite{pointertransformer}, is based on attention and was initially applied to LSTM (Long Short-Term Memory) architectures to address the challenge of generating rare and out-of-vocabulary (OOV) tokens. Inspired by this work, we extend the idea from the LSTM to the Transformer architecture, resulting in a copy-enhanced Transformer specifically tailored for the program slicing task. The overall architecture is illustrated in Fig.~\ref{fig:framework}. Unlike LSTMs, which rely on both the encoder and decoder for recurrence, the Transformer operates purely through attention mechanisms.

In our design, we use cross-attention weights to estimate the copy probability of source tokens. Specifically, we compute: 
\begin{align}
\alpha = \text{softmax}\left(\frac{Q_{dec}K_{enc}^T}{\sqrt{d}}\right)
\end{align}
where $Q_{\text{dec}}$ and $K_{\text{enc}}$ are the query and key matrices of the decoder and encoder, and $d$ is the hidden dimensionality. The attention $\alpha$ signals the decoder to focus on important source tokens. Next, the decoder hidden state $h^*$ of the last layer is computed as the weighted sum of input representations as: 
\begin{align}
h^* =  \alpha \cdot V_{enc}
\end{align}
where $V_{enc}$ represents the value matrix from the encoder. The context vector $h^*$ serves as a dynamic summary of the relevant input content at each decoding step. To decide whether to copy or generate a token, we concatenate the context vector $h^*$ with the decoder input $x_{dec}$, and pass it through a copy layer to calculate the generation probability $p_{gen} \in [0,1]$:
\begin{align}
    p_{gen} &= Sigmoid(W_{gen} \cdot [h^*; x_{dec}] +b_{gen})
\end{align}
where $W_{gen}$ and $b_{gen}$ are learnable parameters of the linear copy module. Here, $p_{gen}$ corresponds to the probability of generating tokens from the vocabulary, while $p_{gen} = (1-p_{gen})$ corresponds to the probability of copying a token from the source code snippet using the attention distribution $\alpha$.

Finally, the model interpolates between the generation distribution $P_{vocab}$ and the copy distribution $P_{\text{copy}}$, producing the final output distribution as:

\begin{align}
P(y) = p_{gen}P_{vocab}(y)+(1-p_{gen})\alpha
\end{align}

This copy mechanism not only directs the decoder’s attention to important source tokens, but also enables the model to directly copy tokens from the code snippet, improving its handling of rare tokens in program slice.

During training, we use the Cross-Entropy Loss to maximize the likelihood of the target sequence. The loss function is defined as:
\begin{equation}
\mathcal{L} = - \sum_{i} p^*_i \log(p_i)
\end{equation}

\noindent where $p^*$ is the distribution of the ground-truth slice, and $p$ is the predicted slice distribution produced by \ourtool.


Additionally, to enhance control over the generation process in the task-specific fine-tuned language model, we prepend special markers during fine-tuning, following prior works~\cite{condition,condition2,grammert5}, to improve performance on downstream tasks. These special tokens serve as explicit signals to guide the model toward generating structured outputs for program slicing. Specifically, we introduce the following markers, 
\textit{<line\_number>},\textit{ </line\_number>}, 
\textit{<code>}, \textit{</code>}, \textit{<criterion>}, \textit{</criterion>}, \textit{<slice>}, and \textit{</slice>}, to help the LM better recognize and process slicing-related components.

\begin{algorithm}[h]
\caption{Constrained Beam Search with Lexical Constraint and Syntactic Constraint}
\label{alg:constrained_beam}

\footnotesize
\SetKwInOut{KwIn}{Input}
\SetKwInOut{KwOut}{Output}
\SetKwFunction{GetAllowedTokens}{GetAllowedTokens}
\SetKwFunction{ApplyMask}{ApplyMask}
\SetKwFunction{NextTokenScores}{NextTokenScores}
\SetKwFunction{TopK}{TopK}
\SetKwFunction{TSED}{TSED}
\SetKwFunction{IsEOS}{IsEOS}

\KwIn{Input sequence $\bm{x}$; Language model $\mathcal{M}$; Vocabulary $\mathcal{V}$; Beam size $K$; Max decoding length $L$}
\KwOut{Program slice $\mathcal{Y}$} 

\tcp{Determine lexically allowed tokens}
$\mathcal{A} \leftarrow \GetAllowedTokens(\bm{x})$

\tcp{Initialize beam with empty sequence}
$\mathcal{B} \leftarrow \{(\bm{y} = [], s = 0)\}$

\For{$t = 1$ \KwTo $L$}{
    $\mathcal{B}_{\text{next}} \leftarrow \emptyset$
    
    \ForEach{beam $(\bm{y}, s) \in \mathcal{B}$}{
        \tcp{Apply Lexical Constraint}
        $mask \leftarrow \ApplyMask(\mathcal{A}, \mathcal{V})$\\
        $\bm{p} \leftarrow \NextTokenScores(\mathcal{M}, \bm{y}, mask)$\\
        $\{(z_k, p_k)\}_{k=1}^K \leftarrow \TopK(\bm{p}, K)$
        
        \tcp{Expand each beam}
        \For{$k = 1$ \KwTo $K$}{
            \tcp{Apply Syntactic Constraint}
            $\bm{y}' \leftarrow \bm{y} \mathbin{\Vert} z_k$\\
            $t_{\text{prev}} \leftarrow \TSED(\bm{x}, \bm{y})$\\
            $t_{\text{cur}} \leftarrow \TSED(\bm{x}, \bm{y}')$\\
            
            \tcp{Skip if syntactic error occurs (TSED $\downarrow$) or end of sequence}
            \If{$t_{\text{cur}} < t_{\text{prev}}$ \textbf{or} \IsEOS($z_k$)}{
                \textbf{continue} 
            }

            $s' \leftarrow s + \log(p_k)$\\
            $\mathcal{B}_{\text{next}} \leftarrow \mathcal{B}_{\text{next}} \cup \{(\bm{y}', s')\}$
        }
    }
    \tcp{Keep top-$K$ beams}
    $\mathcal{B} \leftarrow \TopK(\mathcal{B}_{\text{next}}, K)$
}
\tcp{Return the top-1 output sequence with the highest score}
\KwRet{$\mathcal{Y} \leftarrow \arg\max_{(\bm{y}, s) \in \mathcal{B}} s$}
\label{alg:beam-lex-tsed}
\end{algorithm}

\subsection{Constrained Decoding}\label{sec:decoding}

We apply two novel constraints during decoding stage by modifying beam search and present our algorithm in Algorithm~\ref{alg:constrained_beam}. For each beam search step, we first apply lexical constraint to only allow the token appearing in the original code snippet $\bm{x}$ to be sampled (Line 8 - 11). Next, for each selected candidate, we apply syntactic constraint and filter out the ones that violate syntactic constraint (Line 14 - 21). Below, we introduce each constraint.

\subsubsection{Lexical Constraint}


When a model is fine-tuned to generate code, such as in the task of program slicing, it may produce hallucinated or anti-factual tokens that do not appear in the input snippet, as illustrated in Figure~\ref{fig:motivation}. This issue stems from the fact that language models operate over an unconstrained output space. For instance, at each decoding step, CodeT5(+) samples from a fixed vocabulary of 32,100 tokens based on their probability~\cite{codet5+}, regardless of whether these tokens are relevant or even present in the input. However, program slicing is inherently an extractive code transformation task, where the output must consist only  of tokens drawn from the original code snippet. This mismatch between the model's generative nature and the extractive demands of the task often leads to incorrect slices.

To mitigate this, we introduce a lexical-constrained decoding mechanism that restricts the model to only generate tokens from the input sequence. The core idea is simple: tokens that do not appear in the input should be assigned $zero$ probability during decoding. Specifically, we assign a logit score of $-\infty$ to all disallowed tokens, thereby ensuring their softmax probabilities are zero (Lines 9 - 10 in Algorithm~\ref{alg:constrained_beam}). This mechanism enables \ourtool to generate only valid tokesn, reject unknown identifiers, and catch subtle errors such as spelling mismatches, resulting in more accurate code generation. 


\subsubsection{Syntactic Constraint}


\begin{figure}
  \centering
    {\includegraphics[width=1\linewidth]{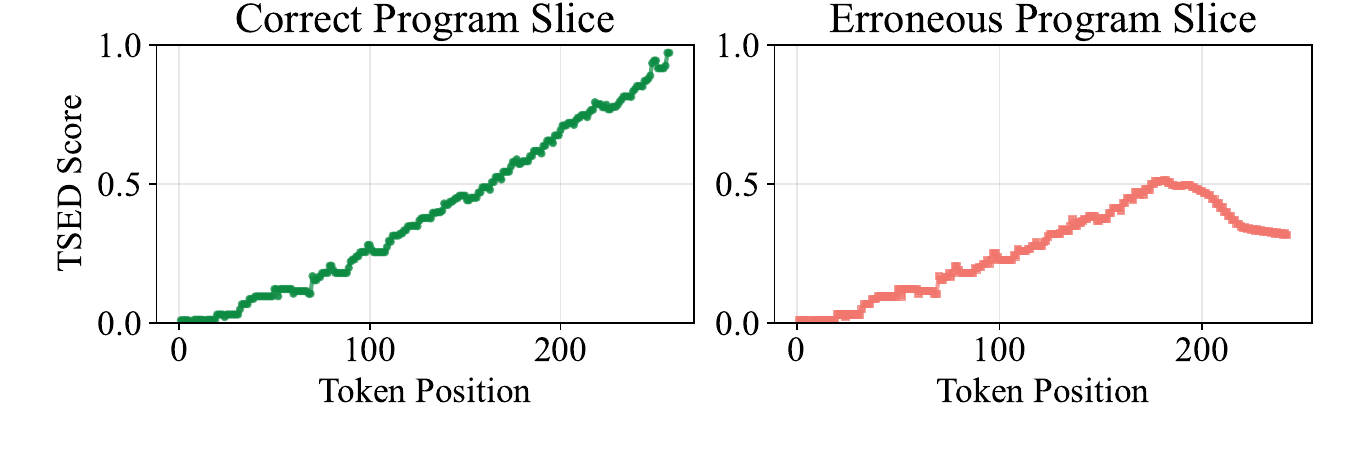}}
    \caption{TSED scores for syntactically correct versus erroneous program slices based on the same input code snippet. The scores are derived from Example (3) in Figure \ref{fig:motivation}. When a syntactic error is introduced into the generated slice, the TSED score deviates from its expected monotonic increasing pattern and instead decreases.}\label{fig:mono}
    \vspace{-0.3cm}
\end{figure}

Lexical constraint alone is insufficient for generating code element preserved code, as they are order-insensitive and do not account for the structure of the output code. For example, as shown in Example (3) in Figure~\ref{fig:motivation}, we observe instances of over-generation \cite{over} in the output code slices. 
Despite the presence of a clear error, all the tokens in the generation are drawn from the input, and thus satisfy the lexical constraint. In such cases, the individual lexical constraint fails to prevent the model generating wrong code because it does not enforce correct token ordering or semantic intent. 

To address this issue, we incorporate an additional AST-based syntactic constraint, namely the Tree Similarity of Edit Distance (TSED)~\cite{tsed} Monotonicity Increasing Constraint. TSED is a structural similarity metric that measures the syntactic distance between two code snippets based on their abstract syntax trees (ASTs). It is defined as: 
\begin{align}
TSED(T_x, T_y) = 1 - \frac{\min_{\text{ops}} \sum_{i=1}^{n} w(op_i)}{\max(\text{Nodes}(T_x, T_y))}\label{fml:tsed}
\end{align}%
\noindent where $\textit{ops}$ denotes the sequence of $n$ edit operations that transform the tree $T_x$ (i.e., the parsed tree of the original code snippet) into another tree $T_y$ (i.e., the parsed tree of the (partial) slice), and $w(op_i)$ represents the cost associated with the $i$-th operation. The computed distance is normalized by the maximum number of nodes in the two trees to account for variations in code complexity and size.

To ensure syntactic coherence during generation, we enforce a monotonicity constraint on the TSED score, requiring that the syntactic similarity between the generated slice and the original code increases monotonically at each decoding step. Since every valid slice is a subsequence of the input code, auto-regressive decoding should incrementally produce outputs that increasingly resemble the original code’s AST, resulting in a progressively higher TSED score. As illustrated in Figure~\ref{fig:mono}, if no structural error occurs, the TSED between the original code and the generated slice should increase monotonically. However, when errors such as over-generation occur, they can disrupt structural similarity and cause a drop in the TSED score. In such cases, the generation process is terminated early, and the output is discarded, as shown in Lines 14–21 of Algorithm~\ref{alg:constrained_beam}.

\begin{figure}
  \centering
    {\includegraphics[width=1\linewidth]{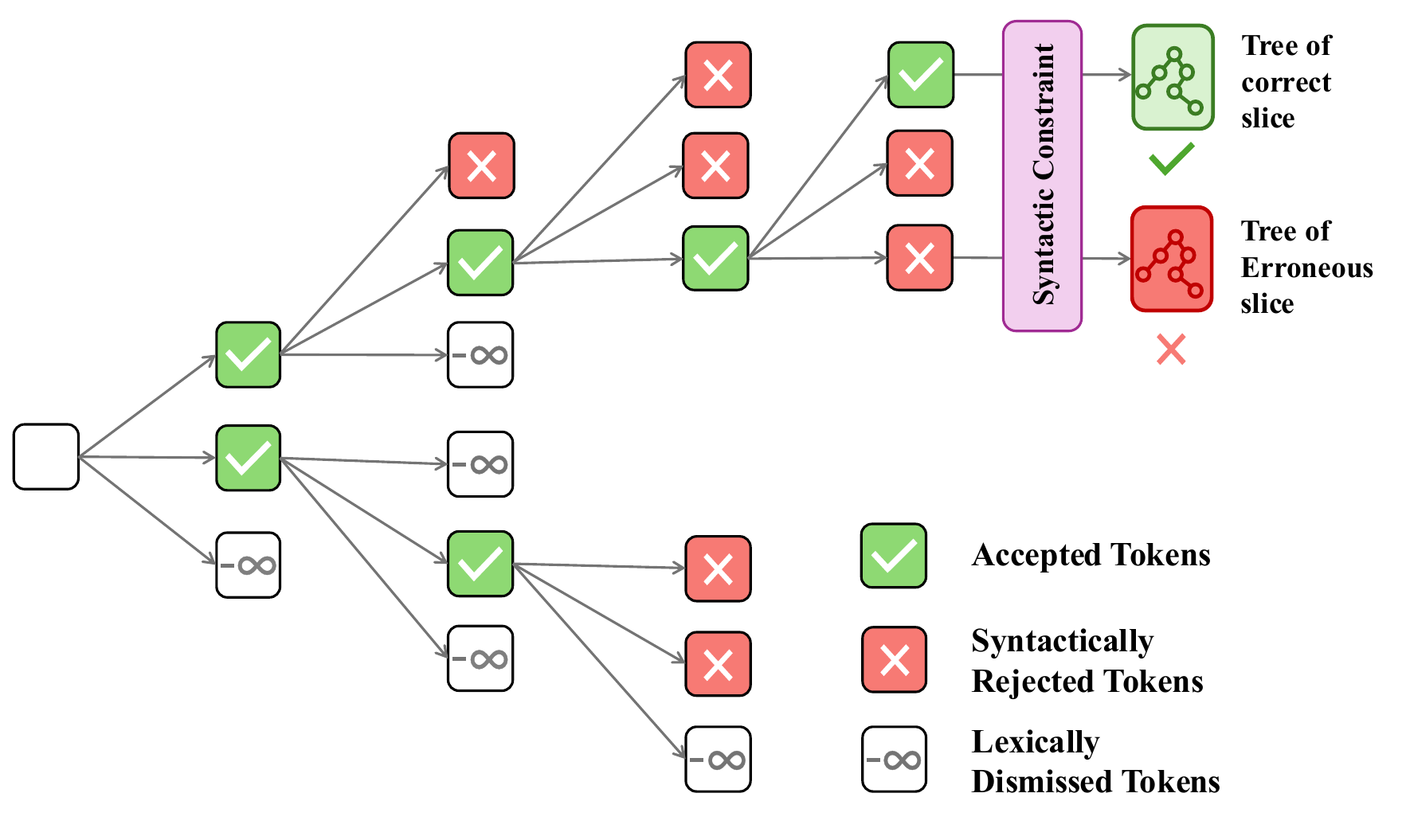}}
    \caption{Implementation of proposed constrained decoding with a beam size of 3.}\label{fig:decoding}
    \vspace{-0.3cm}
\end{figure}

Figure~\ref{fig:decoding} presents the workflow of our constrained decoding process with a beam size of 3. Each vertical column represents the top three token predictions for a hypothesis, arranged in descending order of probability from top to bottom. For each beam search step, the lexical constraint is applied so that only the tokens from the original code snippet and special tokens are allowed to sampled, while all others are automatically dismissed by assigning them a score of $-\infty$. Second, syntactic constraint is applied to measure the TSED monotonicity for each candidate in the beam. Accepted candidates continue decoding in the next step, while rejected ones are terminated early. In the end, we output the top candidate with their highest score. This order-sensitive syntactic constraint complements the lexical constraint by encouraging the model to produce slices that adhere not only to the token-level content but also to the syntactic structure of the input code, thereby significantly reducing the risk of generating ill-formed or structurally implausible slices.

Note that the proposed lexical and syntactic constrained decoding is entirely training-free, and can be directly, easily, and optionally enabled during inference.

\section{Experimental Setting}\label{sec:experimentalsetting}



\subsection{Research Questions}
\begin{itemize}
    \item RQ1: How effective is \ourtool compared to existing SOTA learning-based slicing approaches?
    \item RQ2: How effective is each component of \ourtool?
    \item RQ3: How effective is \ourtool on incomplete code snippets?
\end{itemize}

In RQ1, we compare the effectiveness of \ourtool with prior state-of-the-art (SOTA) learning-based approaches. RQ2 conducts ablation analysis to examine the contribution of each component (i.e., copy mechanism, lexical constraint, and syntactic constraint). In RQ3, we evaluate the effectiveness of \ourtool on incomplete code snippets.

\subsection{Baselines}\label{baselines}

We compare our approach with the following SOTA learning-based baselines.
\begin{enumerate}
    \item \textbf{FM-based slicer~\cite{llmslicer}} leverages foundation models to conduct program slicing~\cite{llmslicer} by using various prompting engineering. Follow their approach, we use three prompting strategies, Zero-shot, Retrieval-augmented Generation (RAG) \cite{allyouneed}, and Chain-of-Thought (COT) \cite{cot}. We select two SOTA foundation language models, GPT-4o-mini \cite{gpt4} and Gemma-7B \cite{gemma2},as the base model.
    \item \textbf{NS-slicer~\cite{learning}} formulates the program slice task as a 0-1 classification.  It applies a CodeBERT \cite{codebert} and GraphCodeBert \cite{graphcodebert} to learning representation then compute the distance between individual statement and criterion. Given an code snippet and slicing criteria, it predicts whether each statement in the code snippet should be included in the slicing or not. 
    \item \textbf{Fine-tune.} To properly assess \ourtool's effectiveness, we compare against a directly fine-tuned model variant that uses the same architecture but without our proposed enhancements (i.e., copy mechanism and constrained decoding). This baseline helps isolate the contribution of our novel components.
\end{enumerate}

\subsection{Datasets and Metrics}
We evaluated the performance of \ourtool on two program slicing datasets: CodeNet \cite{learning} and LeetCode \cite{llmslicer}, following the experimental setups of prior studies~\cite{learning,llmslicer}. The CodeNet dataset consists of programming solutions from IBM, while the LeetCode dataset contains a diverse collection of solutions to various LeetCode problems. Compared to CodeNet, the LeetCode dataset is more complex, featuring longer code samples with more tokens and lines. For both datasets, the ground-truth static program slices were obtained using JavaSlicer~\cite{javaslicer}. In line with prior work~\cite{llmslicer}, we focus exclusively on Java and backward slicing, which is particularly important for applications of static program slicing such as debugging \cite{xu2005brief}. Detailed statistics of the two datasets are provided in Table~\ref{tab:dataset}. We use the training and validation splits of CodeNet to train \ourtool, the test set of CodeNet to evaluate in-domain performance of ours, and the LeetCode test-only set to assess out-of-domain generalization capability.

\begin{table}[ht]
\caption{Basic statistics of our two datasets.}\label{tab:dataset}
\begin{tabular}{l|ccc|c}

\hline
            & \multicolumn{3}{c|}{\textbf{CodeNet}} & \textbf{Leetcode} \\ 
            & train    & valid    & test   & test-only     \\ \hline
Entries     & 30.8K    & 3.5K    & 8.7K   & 100      \\
Avg. tokens & 64       & 64       & 66     & 153      \\
Avg. slocs   & 19       & 18       & 19     & 35       \\ \hline
\end{tabular}
\end{table}

Following previous studies~\cite{learning,llmslicer}, we evaluate the performance of \ourtool on four metrics: 1) \textbf{Dependence Accuracy} (Accuracy-D), measures how accurately statement-level dependencies are predicted. Specifically, Accuracy-D is computed as the ratio of correctly predicted statements to the total number of actual dependent statements for a given slicing criterion in the program.  2) \textbf{Exact Match} calculates the percentage of test instances where the generated program slice exactly matches the ground-truth slice. Even small deviations, such as an incorrect variable name in a statement (e.g., as shown in Example (2) of Figure~\ref{fig:motivation}), are considered failures under this strict metric. 3) \textbf{CodeBlEU}, is a composite metric specifically designed for code generation tasks, CodeBLEU incorporates multiple signals: n-gram matching (BLEU), weighted n-gram match (BLEU-weighted), AST match, and data-flow match~\cite{codebleu}. This metric captures both lexical-level and structural similarities between code snippets. 4) \textbf{TSED} (equation 7), is a novel code similarity metric that compares the abstract syntax trees (ASTs) of the generated and reference code. TSED is sensitive to the syntactic structure of code and is particularly useful for measuring similarity in structured code transformations such as program slicing. Note that NS-Slicer produces only line numbers. To enable a fair comparison using CodeBLEU and TSED, we map the predicted line numbers back to their corresponding program statements before evaluation.

\subsection{Base LMs}

We implemented \ourtool on top of the SOTA seq2seq models CodeT5~\cite{codet5} and CodeT5+~\cite{codet5+}. Those models have been demonstrated to be powerful in code compression and code translation tasks~\cite{codet5,codet5+,li2024few,jiao2023evaluation,yin2024rectifier}. We select the CodeT5-base-0.8B and CodeT5+-0.7B version of those two models, since we wish to keep our approach lightweight. 




\subsection{Implementation Details} 
To ensure consistent outputs, we set the max source and target to 256. The CodeT5(+) is trained using the AdamW optimizer with a batch size of 16, a learning rate of $5 \times 10^{-5}$, and 1,000 warmup steps over 10 epochs. All other settings follow the default CodeT5(+) configuration. We set the beam size to 3,



\section{Results}\label{sec:results}
\subsection{RQ1: Effectiveness of \ourtool}

\begin{table*}[]
\caption{Effectiveness comparison among different static learning-based program slicing methods.}\label{tab:RQ1}
\begin{tabular}{l|cccc|cccc}
\hline
\multicolumn{1}{c|}{\multirow{2}{*}{\textbf{Methods}}} & \multicolumn{4}{c|}{\textbf{CodeNet (in-domain)}}                                                                        & \multicolumn{4}{c}{\textbf{Leetcode (out-of-domain)}}                                                  \\
\multicolumn{1}{c|}{}                                  & \textbf{Acc-D} & \textbf{ExactMatch} & \multicolumn{1}{l}{\textbf{CodeBLEU}} & \textbf{TSED}              & \textbf{Acc-D} & \textbf{ExactMatch} & \textbf{CodeBleu} & \multicolumn{1}{l}{\textbf{TSED}} \\ \hline
GPT-4 (Zero-shot)                                      & 14.76             & 0.00                & 20.10                                 & 35.53                      & 30.91             & 0.00                & 23.77             & 38.69                             \\
GPT-4 (RAG)                                           & 51.70              & 0.00                & 65.18                                 & 59.93                      & 54.09             & 0.00                & 47.05             & 55.41                             \\
GPT-4 (COT)                                            & 56.84             & 0.00                & 68.41                                 & 59.68                      & 60.84             & 7.00                & 46.31             & 54.16                             \\
Gemma (Zero-shot)                                    & 15.23             & 0.00                & 23.30                                 & 43.23 & 33.41             & 0.00                & 22.67             & 42.59                             \\
Gemma (RAG)                                         & 59.23             & 0.00                & 63.24                                 & 59.21 & 41.90             & 0.00                & 38.80             & 42.11                             \\
Gemma (COT)                                          & 64.23             & 0.00                & 74.48                                 & 62.11                      & 41.49             & 0.00                & 40.34             & 48.32                             \\ \hline
NS-slicer (CodeBERT)                                   & 95.65             & 81.72               & 88.41                                 & 91.00                     & 66.43             & 11.00               & 54.91             & 55.60                             \\
NS-slicer (GraphBERT)                                  & 96.51             & 85.77               & 89.26                                 & 90.35                      & 67.07             & 4.00                & 55.45             & 56.45                             \\ \hline
Fine-tune (CodeT5)                                    & 92.19             & 82.80              & 82.74                                 & 82.65                      & 61.34             & 4.00                & 51.38             & 47.40                             \\
\ourtool (CodeT5)                                 & 96.97             & 85.12               & 89.35                                 & 93.95                      & 66.00             & 13.00               & 60.26             & 58.90                             \\ \hline
Fine-tune (CodeT5+)                                   & 95.33    & 87.24     & 89.26                                & 93.42                      & 66.94    & 7.00                & 53.76             & 50.74                             \\

\ourtool (CodeT5+)                                   & \textbf{98.32}             & \textbf{91.30}                & \textbf{92.31}                                 & \textbf{97.06}                      & \textbf{70.85}             & \textbf{14.00}               & \textbf{60.40}             & \textbf{59.97}                             \\ \hline
\end{tabular}
\end{table*}

\textbf{\ourtool consistently outperforms baselines across two datasets and four evaluation metrics. Specifically, \ourtool surpasses the best-performing baseline, NS-slicer, by 6.4\% and 27\% in ExactMatch on CodeNet and LeetCode, respectively.} Table~\ref{tab:RQ1} compares the effectiveness of \ourtool against the evaluated learning-based program slicing methods across four evaluation metrics. \ourtool consistently outperforms all baselines across all datasets and metrics. For example, \ourtool with CodeT5+ achieves 98.32\%, 91.30\%, 92.31\%, and 97.06\% in terms of Acc-D, ExactMatch, CodeBLEU, and TSED on CodeNet, respectively. In comparison, the best-performing baseline, NS-slicer (GraphBERT), achieves 96.51\%, 85.77\%, 89.26\%, and 90.35\% on the same metrics. \ourtool achieves improvements of 1.9\%, 6.4\%, 3.4\%, and 7.4\% on the four metrics, respectively. Even with a less powerful model CodeT5, \ourtool still can achieve the SOTA performance. The main drawback of the previous SOTA, NS-slicer, lies in its design of task modeling.
For example, in Example (4) shown in Figure. \ref{fig:errornsslicer}, NS-slicer models program slicing as a binary classification task, with a fixed threshold to determine whether a statement belongs in the slice. As it embeds each statement independently, it fails to differentiate between identical at different position. Consider the statement (e.g., \texttt{ch = true;}) appearing in two different branches of a method. NS-slicer treats both occurrences identically, even if only one is relevant. In contrast, \ourtool operates at the method level and leverages the full context, allowing it to accurately determine which occurrence is the accurate slice.
\begin{figure}
\begin{tcolorbox}
[boxsep=1pt,left=1pt,right=1pt,top=1pt,bottom=1pt]
\begin{lstlisting}[language=Java,frame=single,framerule=0pt]
// Example (4) - Erroneous slice from NS-slicer
// Expected slice:
...
19: if (Math.abs(as[l]) >= Math.abs(as[r])) {
20: lst.unset(r);
21: ch = true;
...
// Generated slice:
...
20: lst.unset(r);
21: ch = true;
22: }
23   (*@\textcolor{red}{ if (Math.abs(as[r]) >= Math.abs(as[l])) \{}@*)
24: (*@\textcolor{red}{lst.unset(l);}@*)
25: (*@\textcolor{red}{ch = true;}@*)
...
\end{lstlisting}
\end{tcolorbox}

\caption{An erroneous example predicted by NS-slicer.}
\label{fig:errornsslicer}
\end{figure}

\textbf{\ourtool excels at generating fully correct slice, while FM-based method suffers an ExactMatch of 0.} Notably, \ourtool achieves the most significant improvement on the ExactMatch, suggesting that it more frequently generates fully correct slices compared to the baselines, a particularly valuable feature in practical scenarios.
In contrast, the ExactMatch scores for the FM-based methods are nearly all zero. In fact, FM-based methods yield the weakest performance overall. Although techniques like CoT and RAG lead to improvements compared to the zero-shot setting as the results indicate. 
Manual inspection of failure cases reveals that advanced foundation models often misinterpret control flow structures. For example, they may include only the \texttt{if} branch while omitting necessary \texttt{else} or \texttt{else if} branches, leading to logically incomplete slices. In contrast, \ourtool effectively captures these dependencies, allowing it to maintain logical and syntactic correctness in the slices it generates.

\textbf{Compared to the fine-tuned vanilla model, \ourtool achieves significant performance gains, especially in out-of-domain datasets.} As shown in Table~\ref{tab:RQ1}, \ourtool consistently outperforms the fine-tuned orginal CodeT5(+) across on both datasets. For the in-domain dataset (CodeNet), \ourtool (CodeT5+) improves the ExactMatch from 87.24\% to 91.30\%, showing a relative improvement of 4.7\%. Notably, the performance improvement is more pronounced in the out-of-domain dataset (Leetcode), where \ourtool (CodeT5+) doubles the ExactMatch from 7\% to 14\%. We observe a more notable improvement with \ourtool (CodeT5), where the ExactMatch rate is improved from 4\% to 13\% on Leetcode. The results indicate that the enhanced copy mechanism, along with the constrained decoding by lexical and syntactic knowledge, plays a critical role in improving both the slicing accuracy and the preservation of relevant program elements.

\subsection{RQ2: Ablation Analysis}\label{sec:rq2}

\begin{figure}
\centering
\subfigure[CodeNet Dataset]
    {\includegraphics[width=1\linewidth]{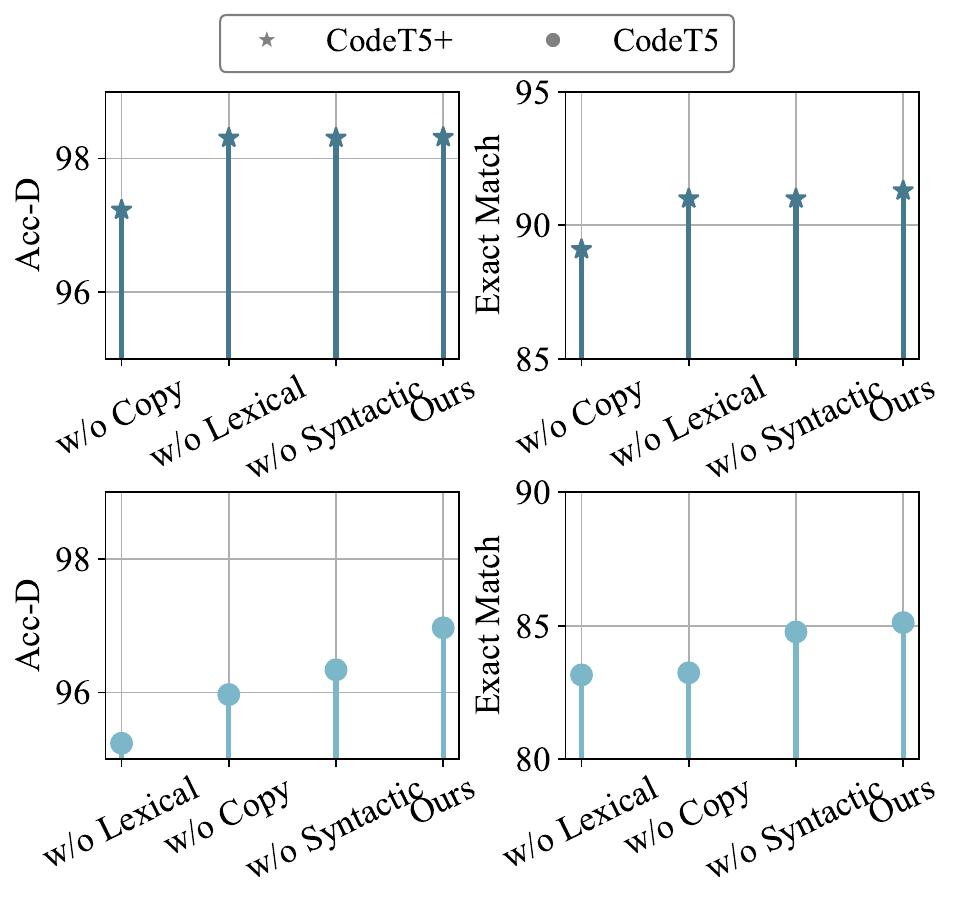}}
       \vspace{-0.5cm}
\subfigure[Leetcode Dataset]
    {\includegraphics[width=1\linewidth]{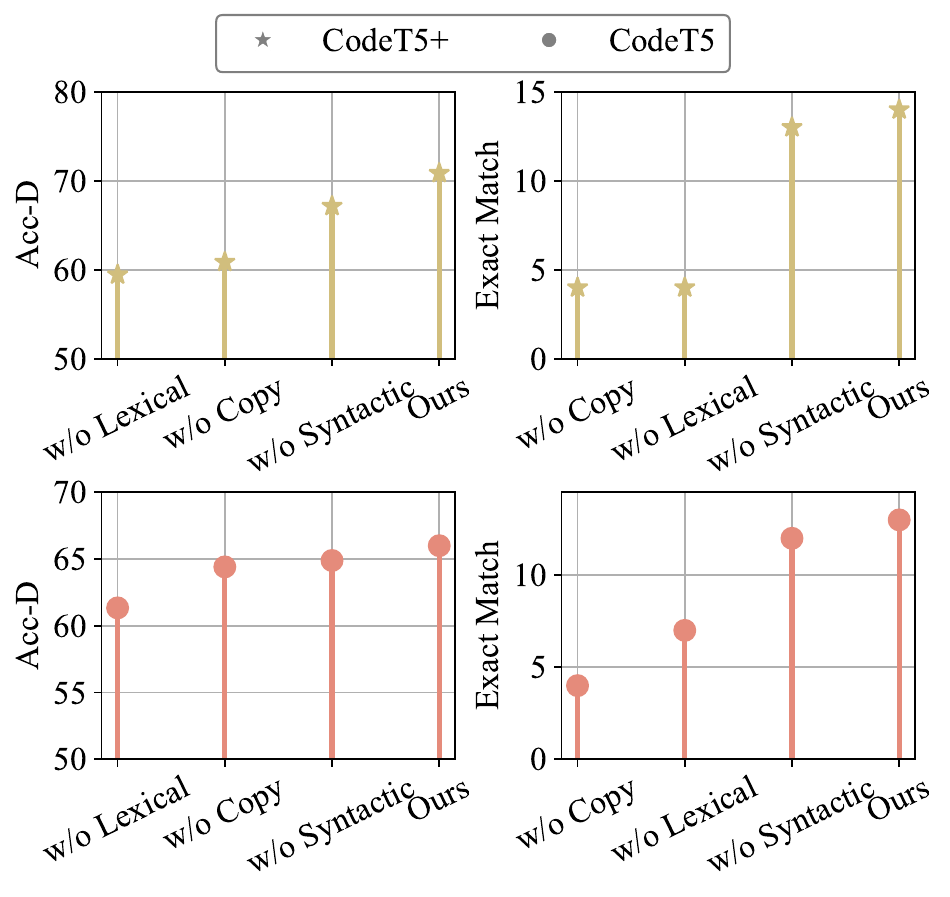}}
    \caption{The ablation analysis results highlight the contribution of each component. A larger performance drop indicates greater importance of the corresponding component. We rank the component importance from left to right.}\label{fig:RQ2}
    \vspace{-0.5cm}
\end{figure}

\textbf{All components make significant contribution to \ourtool.}
Figure~\ref{fig:RQ2} presents the results of the ablation analysis of \ourtool across both base LMs. Overall, all components contribute to \ourtool's performance. As shown, the full version of \ourtool consistently outperforms any variant lacking a single component (i.e., copy mechanism, lexical constraint, or syntactic constraint). Among these components, the copy mechanism and lexical constraint are the most impactful components, with varying influence across metrics. Lexical constraint is key for Acc-D, leading in 3 of 4 LM-dataset settings. For ExactMatch, both components show strong effects, with the copy mechanism also leading in 3 of 4 settings.

\textbf{On out-of-domain dataset, the copy mechanism and the lexical constraint help generate a fully correct slicer remarkably.} When testing on the out-of-domain dataset (Leetcode), we observe that without the copy mechanism and the lexical constraint, ExactMatch drops from 14\% to 4\% and 7\%, respectively. This observation suggests that copy mechanism and lexical constraint make great contribution to ensure generate exactly correct slice.  
The syntactic constraint is the least impactful, likely because structural errors are less frequent than lexical ones, which can be mitigated by the copy mechanism and lexical constraint.


\subsection{RQ3: Effectiveness of \ourtool on Incomplete Code Snippets}\label{sec:rq3}
Compared to program analysis (PA)-based slicing tools such as Javaslicer \cite{javaslicerpaper}, learning-based approaches offer a distinct advantage, which is treating code snippets as token sequences rather than relying on strict syntactic structures. They can be applied more flexibly to incomplete or ill-formed code. To assess the effectiveness of our approach under such conditions, we simulate incompleteness by deliberately corrupting input code snippets before feeding them into the model. Specifically, we evaluate three types of unparsable code, following the setup in \cite{unparsable,errors}: \textbf{1. Missing Class Encapsulation:}
Removing class or method wrappers from the code produces snippets that are not valid compilation units. 
\textbf{2. Missing Semicolons:}
Eliminating semicolons from the ends of statements, which typically leads to parse errors. \textbf{3. Missing or Unmatched Braces:}
Removing required braces or introducing unmatched braces to break block structures and nesting rules.

\begin{table*}[]
\caption{Results of different learning-based methods on the three types of incomplete code snippets. 
}\label{tab:rq3}
\begin{tabular}{lcccccccc}
\hline
\multicolumn{1}{l|}{}                          & \multicolumn{4}{c|}{\textbf{CodeNet}}                                                                                                                                      & \multicolumn{4}{c}{\textbf{Leetcode}}                                                                                               \\
\multicolumn{1}{l|}{\multirow{-2}{*}{Methods}} & \textbf{Acc-D}                    & \textbf{ExactMatch}              & \textbf{CodeBLEU}                  & \multicolumn{1}{c|}{\textbf{TSED}}                            & \textbf{Acc-D}                    & \textbf{ExactMatch}            & \textbf{CodeBLEU}                  & \textbf{TSED}                     \\
\multicolumn{9}{c}{\cellcolor[HTML]{DCDCDC}Missing Class Encapsulation}                                                                                                                                                                                                                                                                                                    \\
\multicolumn{1}{l|}{GPT-4 (Zero-shot)}         & 15.27                     & 0.00                          & 20.14                     & \multicolumn{1}{c|}{36.14}                           & 28.31                      & 0.00                        & 22.43                      & 37.43                      \\
\multicolumn{1}{l|}{GPT-4 (RAG )}              & 53.24                     & 0.00                          & 66.23                      & \multicolumn{1}{c|}{60.05 }                           & 52.40                      & 0.00                        & 46.44                       & 53.89                    \\
\multicolumn{1}{l|}{GPT-4 (COT)}               & 55.45                     & 0.00                          & 67.99                       & \multicolumn{1}{c|}{60.46}                           & 58.93                     & 4.00                     & 44.60                        & 52.19                      \\
\multicolumn{1}{l|}{Gemma 3 (zero-shot)}       & 13.26                     & 0.00                          & 21.20                       & \multicolumn{1}{c|}{38.89}                           & 32.23                     & 0.00                        & 21.24                     & 41.30                      \\
\multicolumn{1}{l|}{Gemma 3 (RAG )}            & 58.27                     & 0.00                          & 61.65                       & \multicolumn{1}{c|}{57.80}                             & 40.15                     & 0.00                        & 37.24                      & 42.55                    \\
\multicolumn{1}{l|}{Gemma 3 (COT)}             & 62.23                    & 0.00                          & 72.28                       & \multicolumn{1}{c|}{59.99}                            & 36.59                    & 0.00                      & 32.35                    & 44.15                    \\ \hline
\multicolumn{1}{l|}{NS-slicer (CodeBERT)}      & 91.61  & 80.96  & 72.53 & \multicolumn{1}{c|}{76.58}                           & 65.32  & 0.00 & 51.47 & 52.25 \\
\multicolumn{1}{l|}{NS-slicer (GraphBERT)}     & 92.10  & 82.10   & 73.08 & \multicolumn{1}{c|}{76.29}                           & 66.59 & 3.00   & 52.65 & 54.31 \\ \hline
\multicolumn{1}{l|}{\ourtool (CodeT5)}             & \textbf{96.37}                      & 89.30                     & 73.96                    & \multicolumn{1}{c|}{78.63}                           & 66.44                     & 13.00                       & 52.75                     & \textbf{54.51}                    \\
\multicolumn{1}{l|}{\ourtool (CodeT5+)}            & 93.30                      & \textbf{91.00 }                       & \textbf{74.34}                      & \multicolumn{1}{c|}{\textbf{80.55}}                           & \textbf{71.73 }                   & \textbf{14.00}                       & \textbf{53.90}                       & 53.33                    \\
\multicolumn{9}{c}{\cellcolor[HTML]{DCDCDC}\textbf{}Missing Semicolons}                                                                                                                                                                                                                                                                                                             \\
\multicolumn{1}{l|}{GPT-4 (Zero-shot)}         & 14.73                     & 0.00                          & 19.67                       & \multicolumn{1}{c|}{35.83}                           & 27.8                      & 0.00                        & 19.89                      & 36.55                     \\
\multicolumn{1}{l|}{GPT-4 (RAG )}              & 52.79                     & 0.00                          & 66.10                      & \multicolumn{1}{c|}{59.89}                            & 51.22                    & 0.00                        & 47.98                    & 52.39                     \\
\multicolumn{1}{l|}{GPT-4 (COT)}               & 55.77                     & 0.00                          & 68.28                     & \multicolumn{1}{c|}{60.26}                           & 59.63                     & 0.00                        & 42.14                      & 51.65                     \\
\multicolumn{1}{l|}{Gemma 3 (zero-shot)}       & 12.48                     & 0.00                          & 22.30                       & \multicolumn{1}{c|}{37.16}                           & 31.20                       & 0.00                        & 22.57                      & 40.99                      \\
\multicolumn{1}{l|}{Gemma 3 (RAG )}            & 55.46                     & 0.00                          & 58.40                      & \multicolumn{1}{c|}{56.40}                             & 43.55                     & 0.00                        & 37.19                     & 41.09                      \\
\multicolumn{1}{l|}{Gemma 3 (COT)}             & 61.52                     & 0.00                          & 71.65                      & \multicolumn{1}{c|}{66.90}                            & 37.68                   & 0.00                        & 36.80                      & 42.85                   \\ \hline
\multicolumn{1}{l|}{NS-slicer (CodeBERT)}      & 95.65                      & 81.72                      & 80.26                       & \multicolumn{1}{c|}{91.00 }                               & 66.43                      & 11.00                       & 54.91                       & 55.60                       \\
\multicolumn{1}{l|}{NS-slicer (GraphBERT)}     & 96.51                      & 85.77                      & 81.41                       & \multicolumn{1}{c|}{90.35 }                            & 67.07                      & 4.00                        & 55.45                       & 56.45                      \\ \hline
\multicolumn{1}{l|}{\ourtool (CodeT5)}             & 97.15                   & 83.80                        & 80.86                      & \multicolumn{1}{c|}{88.94}                            & 67.06                    & 13.00                       & 56.47                      & \textbf{56.75}                      \\
\multicolumn{1}{l|}{\ourtool (CodeT5+)}            & \textbf{97.81}                     & \textbf{89.10}                     & \textbf{84.59}                      & \multicolumn{1}{c|}{\textbf{91.22}}                            & \textbf{69.35}                     & \textbf{13.00}                      & \textbf{56.83}                      & \textbf{55.61 }                     \\
\multicolumn{9}{c}{\cellcolor[HTML]{DCDCDC}Missing or Unmatched Braces}                                                                                                                                                                                                                                                                                                               \\
\multicolumn{1}{l|}{GPT-4 (Zero-shot)}         & 14.70                      & 0.00                          & 19.62                      & \multicolumn{1}{c|}{35.83}                           & 28.17                      & 0.00                        & 21.45                       & 38.14                      \\
\multicolumn{1}{l|}{GPT-4 (RAG )}              & 53.19                    & 0.00                          & 66.62                     & \multicolumn{1}{c|}{60.46} & 53.45                     & 0.00                        & 49.98                     & 55.76                    \\
\multicolumn{1}{l|}{GPT-4 (COT)}               & 55.09                    & 0.00                          & 67.43                     & \multicolumn{1}{c|}{60.14}                           & 61.88                   & 4.00                      & 44.63                       & 57.81                    \\
\multicolumn{1}{l|}{Gemma 3 (zero-shot)}       & 13.53                    & 0.00                          & 23.89                    & \multicolumn{1}{c|}{39.09}                            & 32.63                      & 0.00                        & 19.93                     & 42.32                     \\
\multicolumn{1}{l|}{Gemma 3 (RAG )}            & 56.49                     & 0.00                          & 60.46                       & \multicolumn{1}{c|}{40.51}                           & 46.57                   & 0.00                        & 38.90                       & 41.70                      \\
\multicolumn{1}{l|}{Gemma 3 (COT)}             & 60.10                       & 0.00                          & 73.26                       & \multicolumn{1}{c|}{66.25 }                           & 39.80                     & 0.00                        & 37.25                     & 46.03                   \\ \hline
\multicolumn{1}{l|}{NS-slicer (CodeBERT)}      & 93.42                      & 71.98                     & 78.96                      & \multicolumn{1}{c|}{78.64}                           & 64.31                     & 4.00                      & 55.69                      & 53.66                     \\
\multicolumn{1}{l|}{NS-slicer (GraphBERT)}     & 94.21                    & 72.88                    & 79.27                     & \multicolumn{1}{c|}{79.10}                            & 66.6                       & 4.00                        & 54.76                      & 55.38                    \\ \hline
\multicolumn{1}{l|}{\ourtool (CodeT5)}             & 96.14                    & 73.00                      & 81.57                      & \multicolumn{1}{c|}{81.96}                           & 64.21                     & \textbf{10.00}                     & 56.54                     & 58.00                   \\
\multicolumn{1}{l|}{\ourtool (CodeT5+)}            & \textbf{96.34}                    & \textbf{79.10}                    & \textbf{82.81}                    & \multicolumn{1}{c|}{\textbf{87.75}}                            & \textbf{66.37 }                  & 9.00                    & \textbf{62.19}                    & \textbf{60.28}                   \\ \hline
\end{tabular}
\end{table*}

\textbf{\ourtool consistently outperforms all baselines across all types of unparsable code snippets in terms of all evaluation metrics.} Table~\ref{tab:rq3} compares the performance of different learning-based method on the studied three types of unparsable code snippets. As shown, \ourtool consistently outperforms all other baselines on both datasets across all evaluation metrics. For instance, on Leetcode dataset, \ourtool still achieves the same ExactMatch of 14\%, while the best performing method NS-slicer (GraphBERT)'s ExactMatch degrades from 11\% to 3\%.

\textbf{Among the three types of unparsable code, Unmatched Braces cause the most significant performance drop compared to the other two.} When evaluating all methods studied, including \ourtool, we observe the sharpest decline in performance with Unmatched Braces, as opposed to Missing Semicolons and Missing Class Encapsulation. For example, \ourtool's ExactScore drops to 79.1\% on Unmatched Braces but only to 91.0\% and 89.1\% for the other types. Other learning-based methods exhibit similar trends. These results suggest that Unmatched Braces have the strongest negative impact on learning-based approaches, whereas they remain more robust to missing semicolons or class encapsulation.
One possible explanation is that braces \texttt{\{\}} define block structures (e.g., loops, conditionals, method bodies) and its scope, which is fundamental to program semantics. 
Missing/Unmatched braces break nesting rules, leading to ambiguity in variable scope (e.g., is a variable inside or outside a loop?) and incorrect control flow interpretation (e.g., does an if statement cover the next line or not?). 
Language Models like CodeT5 rely heavily on syntactic structure for downstream tasks such as program slicing, so brace errors severely distort their understanding. In contrast, missing class encapsulation and semicolon removes structure but preserves internal logic and dependency, which leads to less significant impact compared to unmatched braces.


\section{Discussion}
\subsection{Applications of Constrained Decoding in other SE tasks}
In this paper, we proposed a novel constrained decoding strategy that integrate lexical and syntactic constraints. It offers clear advantages that go beyond just program slicing. It can significantly improve the reliability of language models across a range of software engineering tasks by preventing invalid tokens and maintaining structural consistency in the output. 

One practical use case is code completion and automated bug fixing. Lexical constraints help ensure that the model only suggests valid keywords or in-scope variables, which reduces the risk of hallucinated code, which is an issue frequently noted in code generation~\cite{zhang2025llm}. At the same time, syntactic constraints help the model generate code that fits neatly into the surrounding context without introducing syntax errors.
This approach is also valuable for tasks like code refactoring and transformation. By applying constraints, we can ensure that updates follow the correct language conventions and maintain the structure of the original code. This helps reduce the chance of breaking changes during API migrations or modernization efforts.



\subsection{Threats to validity}
\textbf{Threats to external validity} 
In RQ3, we implement and evaluate three types of code corruption to simulate incomplete code snippets. Although, we simulate three most frequent cases on Stack Overflow by following previous studies~\cite{unparsable,errors}, real-world unusable code can be far more diverse in form and severity. We encourage future work to expand the evaluation to additional variants of syntactically incomplete code and assess the robustness of learning-based program slicing methods across these cases.

This study focuses exclusively on Java and two recently released slicing datasets. However, our approach is readily adaptable to other programming languages such as Python, as well as to other extractive coding tasks. We encourage future research to extend this work to additional programming tasks where outputs are subject to structural or semantic constraints. Our experiments include GPT-4o-mini and Gemma-7B as foundation model baselines, though other models may yield better performance. Due to hardware limitations, we only fine-tuned lightweight versions of CodeT5 and CodeT5+ as slicers, and alternative checkpoints may lead to different results.

\textbf{Threats to internal validity} 
Our method is fine-tuned on the full CodeNet slicing dataset, which contains 30.8k examples. As a result, its performance is closely tied to the representativeness and diversity of this training data. Although our approach outperforms baseline methods on the out-of-domain LeetCode dataset, the model still struggles to generalize effectively, achieving a maximum exact match rate of 14\%. This suggests that the current training set may lack sufficient variety. We encourage future work to expand the collection of slicing examples in order to develop more generalizable, learning-based program slicing methods.





\section{Related Work}\label{sec:related}

\subsection{Static Program slicing}


Traditional static slicing tools (e.g., JavaSlicer \cite{javaslicer}, CPP-Slicer \cite{cppslicer}) construct System Dependence Graphs (SDGs) via AST parsing and dependency analysis, then perform slicing as a graph reachability problem \cite{javaslicerpaper}. However, these methods require fully compilable code (at least method-granular) \cite{learning}, limiting their utility for real-world incomplete code snippets like those in Stack Overflow posts \cite{unparsable}.
Recently, learning-based methods have been developed to address this limitation~\cite{learning,llmslicer}. Yadavally et al. \cite{learning} proposed NS-slicer to predict the dependency between slicing criteria (e.g., variable) and statements by analyzing their similarity using CodeBERT/GraphCodeBERT for encoding. However, this approach is not an end-to-end solution. Most notably, it can only utilize the information of a target statement while ignoring the surrounding context, which makes it challenging to process out-of-domain samples. Shahandashti et al. \cite{llmslicer} investigated the application of state-of-the-art large foundation models, such as GPT-4o \cite{gpt4}, GPT-3.5-turbo, and Gemma \cite{gemma2}, to program slicing tasks. Their approach includes leveraging advanced prompting techniques such as retrieval-augmented generation (RAG) and chain-of-thought (COT) reasoning. However, employing foundation models is computationally expensive and yields unsatisfactory results. For instance, in slicing tasks of LeetCode solutions, these models often produce outputs with a zero exact match rate.

Unlike existing learning-based approaches, we proposed a novel lightweight framework that improves existing language model by integrating copy mechanism and constrained decoding. 

\subsection{Pre-trained Language Models for Software Engineering Applications}

Recent advances in pre-trained language models have significantly enhanced capabilities in code understanding and generation. Sequence-to-sequence (seq2seq) architectures, in particular, have emerged as a dominant paradigm, achieving state-of-the-art performance when fine-tuned for various software engineering tasks such as code summarization~\cite{cct5,mastropaolo2024towards,gao2024learning}, translation~\cite{grammert5,yin2024rectifier,zhu2024semi}, and type inference~\cite{typeT5,peng2023generative}. For instance, 
Wei et al.~\cite{typeT5} formulated type inference as a seq2seq code infilling task. They fine-tuned CodeT5 to automatically predict missing type annotations. To enhance the vanilla CodeT5 model, they leverage static analysis to construct dynamic contexts for each code element whose type signature is to be predicted. In addition, they propose an iterative decoding scheme that incorporates previously predicted types into the input context, significantly improving prediction accuracy. 
To better support developers in software maintenance tasks due to code change, such as writing a description for the intention of the code change, or identifying defect-prone code changes, Lin et al.~\cite{cct5} introduced CCT5, a domain-adapted version of CodeT5 specialized for code changes. They collected a large-scale dataset of code diffs and their associated commit messages, and designed five pretraining tasks to capture diverse aspects of code evolution. Zhu et al.~\cite{grammert5} proposed GrammarT5, a grammar-integrated extension of the 
CodeT5 model. Rather than representing code as plain token sequences, GrammarT5 introduces a novel Tokenized Grammar Rule Sequence (TGRS), which embeds syntax information from grammar rule sequences used in syntax-guided generation. To further enrich syntactic understanding, the authors proposed two grammar-aware pretraining objectives, Edge Prediction (EP) and Sub-Tree Prediction (STP) to help differentiate grammar rules across languages. GrammarT5 achieves state-of-the-art performance on a range of code generation, summarization, and translation tasks.

Our work builds on lightweight seq2seq models, CodeT5 and CodeT5+, to the new downstream task static program slicing. Unlike previous efforts, we enhance the original architecture of CodeT5(+) with a copy mechanism and incorporate external lexical and syntactic knowledge by constraining the decoding process.
\section{Conclusion}\label{sec:conclusion}
In conclusion, this paper presents \ourtool, a learning-based approach for static program slicing that leverages a copy-enhanced encoder-decoder Transformer architecture. Our improved model effectively predicts accurate dependency relationships and generates element-preserving program slices. In addition, we introduce a lexically and syntactically constrained decoding strategy, enabling the model to better capture both semantic information and syntactic structure during slicing. We evaluate \ourtool on two program slicing datasets, where it achieves state-of-the-art performance compared to strong baselines, including advanced foundation models and prompting techniques. Specifically, \ourtool outperforms the best baseline by at least 6.4\% and 27\% in terms of ExactMatch on the two datasets, respectively. Furthermore, ablation studies confirm that both the copy mechanism and the constrained decoding strategy contribute significantly to the overall performance improvements.

\clearpage
\bibliographystyle{ACM-Reference-Format}

\bibliography{main}



\end{document}